\documentclass[oneside,fleqn,10pt]{SelfArx} 

\usepackage{afterpage}

\definecolor{color1}{RGB}{0,0,90} 
\definecolor{color2}{RGB}{0,20,20} 
\usepackage{soul}

\usepackage{hyperref} 
\hypersetup{hidelinks,colorlinks,breaklinks=true,urlcolor=color2,citecolor=color1,linkcolor=color1,bookmarksopen=false,pdftitle={Title},pdfauthor={Author}}

\usepackage{graphicx}
\usepackage{subcaption}
\usepackage{float}
\JournalInfo{} 
\Archive{Draft Version} 

\PaperTitle{Studying Illustrations in Manuscripts: An Efficient Deep-Learning Approach} 

\Authors{Yoav Evron\textsuperscript{1}, Michal Bar-Asher Siegal\textsuperscript{2}, Michael Fire\textsuperscript{1}} 

\affiliation{\textsuperscript{1}\textit{Faculty of Computer and Information Science, Ben-Gurion University of the Negev, Be’er Sheva, Israel}} 
\affiliation{\textsuperscript{2}\textit{The Goldstein-Goren Department of Jewish Thought, Ben-Gurion University of the Negev, Be’er Sheva, Israel}} 

\newcommand{\keywordname}{Keywords}

\Abstract{
The recent Artificial Intelligence (AI) revolution has opened transformative possibilities for the humanities, particularly in unlocking the visual-artistic content embedded in historical illuminated manuscripts. While digital archives now offer unprecedented access to these materials, the ability to systematically locate, extract, and analyze illustrations at scale remains a major challenge.

We present a general and scalable AI-based pipeline for large-scale visual analysis of illuminated manuscripts. The framework integrates modern deep-learning models for page-level illustration detection, illustration extraction, and multimodal description, enabling scholars to search, cluster, and study visual materials and artistic trends across entire corpora.

We demonstrate the applicability of this approach on large heterogeneous collections, including the Vatican Library and richly illuminated manuscripts such as the Bible of Borso d’Este. The system reveals meaningful visual patterns and cross-manuscript relationships by embedding illustrations into a shared representation space and analyzing their similarity structure (see figure \ref{fig:network}).

By harnessing recent advances in computer vision and vision-language models, our framework enables new forms of large-scale visual scholarship in historical studies, art history, and cultural heritage making it possible to explore iconography, stylistic trends, and cultural connections in ways that were previously impractical.
}

\begin{document}

\flushbottom 

\maketitle 

\section{Introduction} 

Manuscripts from Late Antiquity, the Middle Ages, and the Early Modern period serve as a unique window into human society's cultural, intellectual, and social worlds during pivotal eras of its history~\cite{de1994history}. These manuscripts are not merely physical objects composed of ink and parchment or paper; they are vibrant reflections of their time's knowledge, beliefs, religions, politics, and arts. Through these texts, we can explore the processes of idea creation, dissemination, and reception, often revealing the individuals behind them. Manuscripts illuminate patterns of life and the challenges of past societies, providing a foundation for understanding the origins of modern culture~\cite{wandrey2017jewish,epstein2015skies}.

Illustrations within manuscripts are more than just decorative elements; they are a vital component of the text that enhances its meaning, accessibility, and cultural value~\cite{bovey2002monsters,kogman2004jewish}. In illuminated manuscripts, illustrations often serve liturgical, pedagogical, or commemorative purposes, reinforcing the sacred or intellectual nature of the text~\cite{epstein2011medieval}.

In many cases, illustrations serve as visual storytelling tools, bridging the gap between literacy and comprehension for audiences of varying educational backgrounds. They provide context to complex narratives, clarify textual ambiguities, and offer symbolic interpretations that deepen the reader's engagement with the material~\cite{epstein2015skies}. Additionally, illustrations reflect the aesthetic preferences, artistic trends, and technological capabilities of the societies in which they were created. They are often beautiful or grotesque, surprising or remarkably realistic~\cite{bovey2002monsters}. They can showcase a deep knowledge of plants and animals~\cite{walker2012medieval}, while also displaying whimsy, humor, theology, heresy, and the artist's exceptional talent. They also reveal cultural exchanges, as artistic motifs and techniques have often traveled across regions and traditions, enriching the visual language of different cultures~\cite{kogman2004jewish}. For scholars of religion, art history, and history in general, these visual elements provide insights into social hierarchies, religious practices, and even material culture, offering a multidimensional view of the past.

However, the sheer scale of digital archives presents a significant challenge: manually identifying and cataloging illustrated pages within millions of scanned documents is an overwhelming and often impractical endeavor~\cite{grana2011automatic,mehri2013pixel}. The exponential growth in digitized content, encompassing manuscripts, books, and historical documents from diverse cultural and temporal contexts, has outpaced traditional methods of scholarly analysis~\cite{ruttenberg2012mass}. Sifting through these extensive collections to locate pages with visual illustrations demands substantial time, labor, and expertise, often making it infeasible for individual researchers or collaborative teams.

Institutions like the Library of Congress,\footnote{Library of Congress Digital Collections, accessed March 2025, \url{https://www.loc.gov/collections/}} British Library\footnote{British Library, Digitised Manuscripts, accessed March 2025, \url{https://www.bl.uk/manuscripts}}, and the Bibliothèque nationale de France (BnF)\footnote{Bibliothèque nationale de France, Gallica Digital Library, accessed March 2025, \url{https://gallica.bnf.fr}} have released tens of thousands of high-resolution scans, covering diverse periods, styles, and cultural contexts. This diversity poses challenges for visual pattern recognition~\cite{ziran2023enhancing}, but it also provides fertile ground for computational learning. The scale and variety of this visual data now make it possible to apply computer vision and deep learning techniques to manuscript analysis. Building on this opportunity, our study presents a scalable and generalizable approach for identifying illustrated pages within digitized manuscripts.

While the challenge of locating illustrations in digitized manuscripts has long been recognized~\cite{mehri2013pixel,garz2011using}, existing solutions often fall short when applied at scale. Previous approaches have relied on image segmentation techniques that attempt to label each pixel in an image as belonging to categories such as text, decoration, or marginalia~\cite{monnier2020docextractor}. While segmentation techniques can be precise in theory, they often require substantial pre-processing (such as binarization and de-skewing) and post-processing (such as filtering or assembling bounding regions) to produce usable results~\cite{mehri2013pixel}. These methods also tend to involve long processing times per page~\cite{he2017mask}, since segmentation models must generate predictions at the level of individual pixels rather than only bounding boxes (i.e., millions of pixel predictions compared to tens of box predictions per page). Consequently, in many practical pipelines, object-detection is employed as a preprocessing stage for subsequent segmentation tasks~\cite{lin2014microsoft}. As a result, such approaches are generally unsuitable for use as off-the-shelf tools in high-volume manuscript digitization projects. Our work builds on this foundation but shifts the focus toward lightweight, generalizable, and scalable deep-learning-based detection methods.

Motivated by the need to process millions of scanned pages, we developed a pipeline designed to identify, extract, and describe visual content in large-scale historical collections. 
In the first stage, a convolutional neural network classifies each page as either “illustrated” or “non-illustrated” depending on the presence of illustration in that page. This allows us to discard the majority of empty pages or those that contain only text, and focus only on the relevant visual material. 
In the second stage, an object detection model locates and crops illustrations, including ornate initials, marginalia, and full miniatures. 
Lastly, we use vision-language models to generate rich human-readable textual descriptions of each detected illustration. 
These captions are stored alongside the cropped images in a database, enabling keyword-based or semantic search across extensive collections. This integrated framework allows scholars to perform queries that were previously impossible to execute manually, for example, searching for phrases such as “winged horse” or “angel holding a sword” and instantly retrieving visually relevant fragments from millions of pages.
The illustrations can be further embedded into a shared representation space and linked into an illustration-similarity graph. This additional layer provides a corpus-level view of the visual landscape, surfacing cross-manuscript relationships and clusters of images that share stylistic, iconographic, or compositional features. This graph-based perspective reveals patterns that remain invisible when pages are examined in isolation, opening new pathways for large-scale, data-driven art-historical and cultural-heritage research.

While the proposed framework is designed as a general and modular pipeline, in this study we implemented it using a specific set of algorithms. The page-level classification stage employs an EfficientNet-based\cite{tan2019efficientnet} convolutional network fine-tuned on over 20,000 manually labeled pages and taken from diverse collections spanning centuries, regions, and artistic traditions. For the object detection stage, we used a YOLO architecture\cite{redmon2016you} trained on more than 1,500 annotated examples sourced from various Vatican manuscripts and related materials. Finally, for the captioning stage, we utilized the LLaVA (Large Language and Vision Assistant) vision-language model\cite{li2023llava} to generate detailed textual descriptions for each extracted illustration. We emphasize that the framework itself is modular by design, and as new algorithms emerge, each stage can be independently improved or replaced to enhance overall performance.
In our current implementation, the classification model achieved a ROC-AUC of 0.95, demonstrating strong overall discrimination between illustrated and non-illustrated pages. At the optimal decision threshold, it reached a precision of 78.6\%, recall of 74.6\%, an F1-score of 76.5\%, and an overall accuracy of 95.1\% on the held-out test set.

To evaluate our method in a realistic, large-scale setting, we applied the whole pipeline to more than 10,000 items, including over five million scanned manuscript pages from the digitized collection of the Vatican Library.\footnote{Vatican Apostolic Library, Vatican Digital Library, accessed March 2025, \url{https://digi.vatlib.it/}} These collections span diverse periods, regions, and scripts~\cite{manoni2017vatican}. The dataset includes complex layouts, varying resolutions, and frequent image artifacts such as stains, tears, or marginal notes-characteristics that pose significant challenges for computational analysis~\cite{philips2020historical, kapon2024shapinghistoryadvancedmachine}.

Our classification model filtered out over 90\% of pages classified as text-only, allowing the object detection and captioning stages to focus on a smaller, more relevant subset. Overall, we identified and extracted more than 350,000 unique illustrations. The average processing time was under 0.06 seconds per page, enabling end-to-end analysis of the entire Vatican corpus within days. The illustrations were automatically described using the AI image-captioning model and indexed in a searchable database. Qualitative inspection revealed a wide variety of visual motifs, from elaborate miniatures to marginalia and decorated initials.

Our framework not only automates a historically labor-intensive task, but also enables a new mode of scholarship: for the first time, researchers can systematically compare, retrieve, and analyze visual motifs across vast manuscript collections across time, geography, and artistic traditions. This paradigm shift opens the door to new discoveries in art history and cultural studies, making it possible to detect patterns, symbolic connections, and stylistic trends that have remained hidden in plain sight for centuries.

Our main contributions are as follows:

\begin{itemize}
\item We release an open scalable deep learning pipeline for identifying, extracting, and describing visual content in digitized historical manuscripts.

\item We demonstrate the effectiveness of our two-stage approach across noisy and heterogeneous manuscript pages, which performs reliably across noisy, irregular manuscript pages, with inference time of under 0.1 seconds per page,\footnote{Experiments were conducted on an Intel Core i7-1355U (13th Gen), 16 GB RAM, Intel Iris Xe GPU.} enabling efficient processing of millions of pages.

\item We deploy the extracted visual content in a searchable platform that enables humanities scholars to retrieve illustrations based on semantic queries - available upon request.

\item Our pipeline enables new types of cultural and iconographic inquiries, for instance, tracing the evolution of symbolic motifs, identifying shared visual programs across distant manuscripts and artists, or discovering latent visual connections previously inaccessible to human review.
\end{itemize}

The remainder of this paper is organized as follows:
Section \ref{sec:related} provides an overview of related work in illustration detection and manuscript image analysis. Section \ref{sec:methods} describes our methodology, including dataset construction, model selection, and pipeline implementation. Section \ref{sec:results} presents the evaluation setup and results, including qualitative examples and runtime benchmarks. Section \ref{sec:discussion} outlines potential applications and limitations. Finally, Section \ref{sec:conclusions} summarizes the main findings and discusses future  directions for improving illustration retrieval and understanding in historical archives.


\section{Related Work} 
\label{sec:related}

In this section, we review existing approaches to the analysis of historical manuscript. 
In Section~\ref{subsec:Digitization}, we outline the large-scale digitization of historical collections. In Section~\ref{subsec:IllustrationDetection}, we examine methods for illustration detection, ranging from traditional techniques such as Optical Character Recognition (OCR) and Page Layout Analysis (PLA) to recent computer vision approaches, including image classification and object detection.
In section~\ref{subsec:rel-caption} we provide an overview of advances image captioning techniques and vision-language models and their emerging applications to historical collections. 
Finally, Section~\ref{subsec:rel-embedding} discusses image-embedding methods and graph-based similarity representations.

\subsection{Digitization}
\label{subsec:Digitization}
In recent years, major libraries have substantially expanded their digitization initiatives, making an unprecedented amount of collections of historical manuscripts accessible to scholars~\cite{ruttenberg2012mass}. Institutions such as the British Library, the Bibliothèque nationale de France, and the Library of Congress now offer significant portions of their historical archives online, spanning across diverse time periods, and languages. Gallica, the digital library of the Bibliothèque nationale de France, alone offers access to millions of high-resolution pages~\cite{Duchesneau_2014} while the British Library’s “Digitised Manuscripts” collection features over 8,000 items~\cite{prescott2018we}. This digitization surge has highlighted the lack of scalable tools for analyzing digitized books in general, and visual elements such as illustrations, which remain understudied compared to text~\cite{suissa2022text}.

This abundance of material has significantly reshaped humanities research~\cite{Mughaz2015}. For instance, these manuscripts reveal the cultural dynamics of Jewish communities in their interactions with surrounding societies and how distinct Jewish traditions were formed and shaped during periods of both prosperity and persecution~\cite{kogman-appel_jewish_2004}. Beyond their cultural and historical value, manuscripts constitute the primary source material for constructing diachronic corpora, enabling linguists to investigate orthographic and grammatical changes over time. For example, large-scale historical corpora spanning more than 1,400 years in Arabic have been built from digitized manuscripts, supporting automated periodization and morphological analysis~\cite{belinkov2018studyinghistoryarabiclanguage}. In art history, digitized manuscripts enable comparative studies like the depiction of dragons, angels, or botanical illustrations across regions and time periods~\cite{bovey_monsters_2002}. Another important line of research has examined medical drawings in medieval manuscripts, revealing how anatomical sketches and disease depictions shaped the transmission of medical knowledge across Europe and the Middle East~\cite{mitchell2016anatomy, gurunluoglu2013history, petaros2013anatomical}.

Large-scale digitization necessitates reliable image storage and retrieval solutions. For this purpose, many institutions have adopted the International Image Interoperability Framework (IIIF)~\cite{snydman2015international}. IIIF provides standardized APIs to deliver, annotate, and display high-resolution images across repositories, facilitating interoperability and consistent metadata management. For example, the Vatican Library uses IIIF to provide access to millions of digitized pages from tens of thousands of manuscripts~\cite{piazzoni2024process}.

\subsection{Illustration Detection in Historical Manuscripts}
\label{subsec:IllustrationDetection}
A significant portion of historical document analysis techniques focus on textual extraction, particularly through Optical Character Recognition (OCR) and Handwritten Text Recognition (HTR)~\cite{vamvakas2008complete,tsochatzidis2021htr,martinek2020building}. While textual content is effectively being extracted in these approaches, non-textual elements such as illustrations or marginalia are often ignored. Tools such as Transkribus~\cite{kahle2017transkribus} and eScriptorium~\cite{kiessling2019escriptorium}, explicitly developed for historical manuscripts, employ machine learning-based models to recognize handwritten text and perform region or line segmentation. However, these systems are primarily optimized for textual content, and are not designed to detect or analyze visual elements such as illustrations or decorated initials at large scale~\cite{mehri2013pixel, rotman2022detectionmaskingimprovedocr}. Moreover, while these platforms perform well on Latin-script manuscripts, applying them to Semitic languages that are written right-to-left poses distinct challenges due to different writing directions, diacritics, and script structures.

Another common approach is Page Layout Analysis (PLA), which segments pages into distinct content regions (e.g., headings, paragraphs, images)~\cite{sven2022page, kodym2021page, bukhari2012layout}. Although effective for comprehensive document understanding, traditional segmentation based methods such as docExtractor~\cite{monnier2020docextractor}, often rely on pixel-level analysis and require complex pre- and post-processing pipelines~\cite{cheng2020novel, paluspreprocessing}. These techniques can be computationally intensive and may not scale efficiently to large manuscript collections~\cite{iglesias2015multi}.

While layout segmentation tools occasionally identify illustration regions, few methods explicitly aim to extract illustrations as discrete visual units suitable for standalone analysis. This distinction is critical, especially when illustrations are embedded within ornate initials, marginalia, or complex text-visual arrangements. Traditional layout methods often treat such elements as secondary byproducts rather than primary targets~\cite{rotman2022detectionmaskingimprovedocr}. Moreover, detecting illustrations in historical manuscripts poses unique challenges due to wide stylistic variation, irregular spatial organization, and frequent physical degradation (e.g., stains, bleed-through, or torn edges)~\cite{philips2020historical, kapon2024shapinghistoryadvancedmachine}.

Recent years have witnessed rapid progress in deep learning for image classification and object detection~\cite{alom2018historybeganalexnetcomprehensive}. In classification, convolutional neural networks (CNNs) have evolved from early architectures like AlexNet~\cite{krizhevsky2012imagenet} and VGG~\cite{simonyan2015vgg} to more efficient and accurate designs such as EfficientNet~\cite{tan2019efficientnet} and ResNet~\cite{he2016resnet}, which achieve state-of-the-art results~\cite{rawat2017deep}. These models are particularly effective for identifying abstract visual categories under varying conditions and have been widely adopted in domains requiring robust generalization, including historical document classification.
For instance, CNN-based classifiers have been used to distinguish between pages in incunabula containing text, tables, pictures, titles, and handwriting ~\cite{ropel2025unfoldingpastcomprehensivedeep}. Other studies have trained classifiers to detect Arabic manuscripts authors ~\cite{khayyat2020towards}.

Parallel advancements in object detection models have produced architectures that combine high accuracy with real-time performance. Frameworks such as YOLO (“You Only Look Once”)~\cite{redmon2016you}, have demonstrated strong performance across diverse detection tasks~\cite{kang2025object}, including challenging domains like digitized archives~\cite{malashin2025recognition}. For example, YOLO models have been successfully applied to detect text lines and text characters in Ottoman manuscripts~\cite{kutal2025text} and birch-bark manuscripts~\cite{malashin2025recognition}. Their ability to process images quickly - often in several milliseconds per image~\cite{terven2023comprehensive} makes them ideal for large-scale archival applications.

\subsection{Image Captioning and Vision-Language Models}
\label{subsec:rel-caption}

Recent breakthroughs in vision-language modeling have transformed the ability to describe and interpret visual content using pretrained image-to-text models~\cite{vinyals2015show, du2022survey}. Modern image captioning systems integrate visual encoders, often based on vision transformers (ViTs) with large language models (LLMs), enabling the generation of fluent and rich captions for images. Architectures such as BLIP (Bootstrapped Language-Image Pretraining)~\cite{li2022blip}, CLIP~\cite{radford2021learning}, and LLaVA~\cite{li2023llava}, exemplify this approach by pairing an advanced image encoder with a language model that can either generate text or follow instructions in natural language.

In particular, BLIP introduced a framework that combines image-text matching with caption generation through a two-tower transformer design. CLIP showed how contrastive training on large sets of image–text pairs can bring visual and textual features into a shared space, which makes zero-shot classification and cross-modal retrieval possible. LLaVA builds on CLIP’s image encoder together with a LLaMA language model, resulting in strong performance on open-ended tasks like visual question answering and instruction following.

While these models were primarily trained on modern images, several recent works have begun exploring their application to historical content~\cite{thomas2024capturing, cetinic2021towards, gupta2020towards}. For example, Thomas and Testini (2024) investigated the automated identification and analysis of image captions in a large corpus of historical book illustrations~\cite{thomas2024capturing}. Cetinic (2021) proposed methods for generating and evaluating iconographic image captions tailored to artworks~\cite{cetinic2021towards}. Similarly, Gupta et al. (2020) explored approaches for adapting image captioning models to art-historical datasets~\cite{gupta2020towards}.

\subsection{Image Embedding and Graph-Based Similarity}
\label{subsec:rel-embedding}
Image embedding techniques transform visual inputs into compact vector representations that capture semantic and stylistic features\cite{he2016resnet}. This dimensionality reduction significantly reduces storage requirements and computation time, while preserving the most relevant visual information, making embeddings particularly suitable for large-scale analysis and retrieval, and are particularly powerful for identifying relationships between images\cite{bengio2013representation}. Modern image-embedding methods rely primarily on deep neural architectures such as convolutional networks (e.g., ResNet)~\cite{he2016resnet} and vision transformers (ViT)~\cite{dosovitskiy2020vit}. More recent multimodal models, including CLIP~\cite{radford2021learning} and ALIGN~\cite{jia2021scaling}, have also demonstrated remarkable performance across a wide range of visual understanding tasks. In the context of cultural heritage, image embeddings have been commonly used. For example, Garcia et al.\ proposed an embedding framework for art-historical images, learning visual representations that integrate metadata such as artist, period, and school in order to improve tasks like author attribution, style classification, and cross-modal retrieval~\cite{garcia2019contextaware}. 
Similarly, Springstein et al.\ developed the iART system, a large-scale art-historical image search engine that relies on deep learning embeddings to cluster artworks, support similarity-based exploration, and facilitate comparative analysis across collections and periods~\cite{springstein2021iart}.

Several works have proposed constructing similarity graphs in which nodes represent embedded images and edges encode visual proximity~\cite{yamins2014performance}. These graphs support community detection, motif tracking, and corpus-level exploration of visual themes, enabling researchers to trace stylistic shifts and uncover connections across manuscripts. For example, Wan et al.\ used a graph-based re-ranking method for image retrieval, modeling the global structure of visual similarity through a kNN graph to significantly improve retrieval precision~\cite{wang2012manifold}.


\section{Methods}
\label{sec:methods}

Our research addresses the challenge of efficiently retrieving illustrations from digitized historical manuscripts using computational methods that can scale to millions of pages in practical time. To meet this challenge, we developed a modular pipeline composed of three main stages: (1) \textit{Image Extraction}, where a classification deep learning model filters out text-only pages and an object detection deep learning model identifies and crops illustration regions; (2)\textit{ Image Captioning}, where a vision-language model produces rich textual descriptions for each cropped illustration ; and (3) \textit{Search and Retrieval}, where both illustrations and captions are stored in a searchable database, enabling scholars to perform keyword-based queries through a web interface. In addition, we construct image-similarity graphs from the extracted illustrations, enabling corpus-level analysis and the discovery of recurring visual patterns and stylistic trends across manuscripts.

While the framework is designed to be general and modular, capable of integrating different models or algorithms at each stage, in this study we implemented it using a specific configuration: an EfficientNet-based classifier for page-level filtering, a YOLO-based detector for illustration localization, and the LLaVA model for caption generation. This concrete implementation serves as a proof of concept, while the pipeline itself remains adaptable to future advances.

This section outlines the datasets, processing steps, methodologies, and evaluation procedures used to build our system. In Section~\ref{methods-data}, we describe the manuscript page images that serve as the dataset of our study. Sections~\ref{methods-classification} and~\ref{methods-detection} then detail the illustration extraction process, including data preprocessing, annotation, model training, and the metrics used for evaluation in both images classification and object detection stages. Section~\ref{methods-caption} discusses how we applied multimodal vision-language models for caption generation. Section~\ref{methods-search} presents how the resulting illustrations and captions form the backbone of the retrieval system, enabling scholars to efficiently locate visual material through keyword searches. Finally, Section~\ref{methods-similarity} describes how we construct illustration-similarity graphs from the extracted images, supporting corpus-level analysis and the discovery of visual patterns across manuscripts.

\subsection{Data Acquisition}
\label{methods-data}
We collected digitized pages from the Vatican Library’s public IIIF API~\cite{vaticanIIIF}. We likewise incorporated the Bible of Borso d’Este\footnote{Accessed November 2025 via the Italian Digital Library, \url{https://edl.cultura.gov.it/media/schedaopen?id=3015932}}. These platforms provide high-resolution color scans of historical documents. Our project focused on two document types: medieval manuscripts - handwritten documents created before the invention of printing; and incunabula, printed books produced before the year 1501. Both types offer rich visual content and span diverse historical periods, regions, and artistic traditions.

The Vatican Digital Library consists of 96 collections of manuscripts and incunabula from different regions, periods, languages, and artists; together, these collections comprise 28,814 items encompassing 9,664,009 pages~\cite{vaticanIIIF}.\footnote{The reported numbers are accurate as of January 2025, when the data were collected.} 
The Bible of Borso d’Este is a 15th-century Italian Renaissance masterpiece, renowned for its breathtaking illuminations, intricate decorative borders, and finely painted marginal scenes~\cite{BorsoEste}. Often described as the “Mona Lisa” of illuminated manuscripts, this codex exemplifies the pinnacle of artistic achievement in Renaissance book production.

\subsection{Illustration Presence Classification}
\label{methods-classification}
\subsubsection{Data Preparation and Labeling}
We constructed the training dataset for the classification stage through a two-step process. In the first step, we randomly sampled 1,000 images from a diverse range of manuscript and incunabula and manually labeled each page as either "art" or "no-art", depending on the presence of illustrations. To ensure that our models would generalize well across volumes from different geographic regions, artists and time periods, we randomly sampled pages from the full range of manuscript and incunabula volumes available via the IIIF API~\cite{vaticanIIIF}, intentionally selecting pages from different volumes, time periods, and creators. This strategy aimed to expose our models to the broadest possible range of layouts and illustration styles to reflect the heterogeneity of historical sources.

Since the images came from different manuscripts, we applied a uniform naming convention and directory structure to track each image’s source volume, page number, and manuscript metadata throughout the pipeline to ensure that subsequent search and retrieval steps would work seamlessly.

Using this initial 1,000-image dataset, we fine-tuned a preliminary classifier, which was then employed to predict labels for an additional randomly selected set of 20,000 pages. In the second step, these automatically generated predictions were carefully reviewed and manually corrected to ensure label accuracy and consistency. This semi-automated approach provided a practical balance: it enabled a substantial expansion of the labeled dataset while reducing the time and effort required for exhaustive manual annotation.

As a result of this pipeline, we constructed a robust dataset of approximately 20,000 images, each reliably labeled as either "art" (illustrated page) or "no-art" (non-illustrated page). This dataset served as the foundation for training and evaluating our subsequent classification model and object detection one.

\subsubsection{Class Imbalance}
Illustrated pages constituted a significant minority in the full corpus. While only 1,173 of the 20,000 images contained illustrations (5.8\%), the remainder consisted of blank pages or text-only pages. To mitigate this extreme imbalance, we retained all illustrated pages while randomly discarding 9,000 non-illustrated pages. This yielded a working dataset of 11,000 images. This downsampling preserved sufficient variation in the negative class ("no art") while improving the positive ("art") ratio to 10.6\%. While this did not yield a perfectly balanced dataset, it substantially reduced the original skew (from 5.8\% positives to 10.6\%), which we deemed sufficient for stable training. More aggressive balancing techniques-such as oversampling or synthetic data generation-were intentionally avoided, as these can introduce artifacts and distort the natural distribution of the corpus. Instead, we prioritized maintaining the natural diversity of the original corpus while reducing imbalance to a level that the model could effectively learn from. Similar downsampling strategies had been also employed in prior work to manage highly skewed datasets \cite{More2016}. Although hybrid approaches that combine oversampling and downsampling are often reported as yielding the most robust results \cite{johnson2020effects}, we opted for downsampling alone in order to avoid introducing synthetic artifacts and to preserve the natural distribution of the corpus.

\subsubsection{Evaluation Metrics}
We evaluated our classification model on a held-out test set of 1,105 images (around 10\% of the labeled data). Following standard practice, we report Accuracy, Precision, Recall, F1-score, AUC, and PR-AUC as performance metrics. In addition, given the class imbalance inherent to our dataset, F1-score provides a more reliable measure of overall performance than accuracy alone. Since the vast majority of pages do not contain illustrations, false positives would overwhelm downstream applications with irrelevant material such as stained or purely textual pages. For this reason, we place particular emphasis on Recall, ensuring that illustrated pages are reliably identified for subsequent processing.

The dataset was split into 70\% training, 20\% validation, and 10\% testing. This proportion reflects standard practice in deep learning~\cite{chen2024image, minisini2024transfer, he2016image}, allocating the majority of data to training while keeping substantial, disjoint partitions for hyperparameter tuning and final evaluation. After this separation, the training set contained 821 illustrated pages and 6,908 non-illustrated pages.

\subsubsection{EfficientNet Fine-Tuning}
For the classification stage, we fine-tuned an EfficientNet-B0 model~\cite{tan2020efficientnetrethinkingmodelscaling}, selected for its favorable trade-off between accuracy and computational efficiency on high-resolution image classification tasks~\cite{tan2020efficientnetrethinkingmodelscaling, tan2021efficientnetv2smallermodelsfaster}. Leveraging a pre-trained model allowed us to transfer features learned from large-scale natural image data (ImageNet~\cite{deng2009imagenet}) to our domain of manuscript pages, where labeled data is comparatively scarce. 
Although more recent large-scale image datasets have been introduced, ImageNet remains widely adopted, as its visual diversity has been shown to generalize effectively across domains, including non-natural imagery. Given that our dataset of manuscript pages is visually distinct from everyday photographs, initializing the model with ImageNet-pretrained weights provided a stable starting point for effective adaptation to our domain.

Training followed a two-stage fine-tuning protocol. In the first stage, we froze the convolutional backbone and trained only the final classification layer for 10 epochs, using binary cross-entropy loss and the ADAM optimizer with a learning rate of $10^{-3}$ that was chosen empirically, following common transfer-learning practice~\cite{howard2018universal}. In the second stage, we unfroze the top portion of the network (the last 20 layers, corresponding to the final two blocks in EfficientNet-B0) and continued fine-tuning the entire network with a reduced learning rate of $10^{-5}$ that was also chosen empirically, following common transfer-learning practice~\cite{howard2018universal}. Early stopping based on validation loss was employed to mitigate overfitting. To improve generalization, we applied standard data augmentations (random horizontal flips and rotations up to 20°) using TorchVision.\footnote{Implemented via \texttt{torchvision.transforms}.} Input images were resized to $224\times224$ pixels - the standard input resolution for EfficientNet-B0 to ensure direct compatibility with the pre-trained model. The images were then normalized using the mean and standard deviation of the ImageNet dataset.

The final model was trained on 11,000 labeled pages and evaluated on a held-out test set (10\% of the data). Once reliable performance was established, we used the model to automatically classify over five million manuscript pages from more than 10,000 volumes. This large-scale application demonstrates the practicality of our pipeline: by pre-filtering over five million pages from more than 10,000 manuscripts, the system efficiently excluded non-illustrated pages helping reduce the computational cost of illustrations detection and captioning, which highlight the feasibility of applying our method beyond research prototypes to real-world, libraries collections.

\subsection{Illustration Localization - Object Detection}
\label{methods-detection}
After identifying the pages that contain illustrations, the next step was to pinpoint the exact regions of these illustrations. In many cases, illustrations occupy only small portions of a page otherwise filled with text, making it inefficient to store entire pages and lowering the accuracy of subsequent captioning. To address this, we fine-tuned YOLOv11 \cite{redmon2016you, khanam2024yolov11}, chosen for its balance between detection accuracy and computational efficiency, which makes it well suited for processing millions of manuscript pages. While bounding-box localization is sufficiently precise, it's far more practical than pixel-level segmentation for large-scale datasets \cite{lin2015microsoftcococommonobjects, zhang2021segmenting}.

\subsubsection{Data Preparation and Annotation}
We constructed a dedicated annotation dataset consisting of 1,800 pages randomly sampled from those previously flagged as illustrated during the classification stage. In selecting these pages, we prioritized diversity across artists, regions, and styles to maximize the model’s ability to generalize. Each image was manually annotated with bounding boxes using the open-source tool \textit{LabelImg} \cite{labelImg}, marking all instances of visual content, including marginalia, decorative initials, and embedded miniatures. To enable the search and retrieval system to also return sub-illustrations - for example, a main character or an animal within a larger scene - we annotated not only the overall illustration but also its major subcomponents (e.g., a specific character or figure within the drawing).
Three pages containing only noise were excluded.

\subsubsection{Evaluation}
The final dataset was split into 70\% training, 20\% validation, and 10\% testing, following common practice in object detection tasks. Annotations were saved in YOLO format, with bounding box coordinates recorded relative to image dimensions. 
Evaluation was based on mean Average Precision (mAP), precision, and recall, providing complementary insights into localization accuracy and detection robustness, following common practice in object detection tasks.

\subsubsection{YOLOv11 Fine-Tuning}
We fine-tuned the lightweight YOLOv11n architecture, initialized with pre-trained weights, for 50 epochs with a batch size of 16 on a dedicated GPU server.\footnote{The server is equipped with an NVIDIA RTX 6000 Ada, 48GB.} Default data augmentation strategies provided by the \textit{Ultralytics}~\cite{Jocher_Ultralytics_YOLO_2023} implementation were used, including color perturbations, translation, scaling, flipping, and mosaic augmentation. Training followed the default Ultralytics~\cite{Jocher_Ultralytics_YOLO_2023} configuration, which employed the AdamW optimizer with automatic hyperparameter tuning, early stopping to prevent overfitting, and mixed-precision optimization.

Once trained, the detector was deployed on the full corpus of pages identified as containing illustrations. For each page, the model extracted cropped regions corresponding to illustration bounding boxes, producing isolated visual segments that have been stored in organized by volume, collection and library for efficient search and retrieval system.

To contextualize these results, we compared our classifier with \texttt{docExtractor}\cite{monnier2020docextractor}, a segmentation-based system. For an optimal comparison, we converted the pixel-level segmentation outputs of \texttt{docExtractor} into bounding boxes. Specifically, we applied their algorithm to the 179 test images to obtain binary masks of illustration regions, then performed morphological closing to merge nearby areas and removed small noisy regions. Bounding boxes were then extracted from the cleaned masks and evaluated against the YOLO ground truth labels. After experimenting with different values (3, 5, and 10 pixels) for the morphological closing parameter, we found that a 10-pixel tolerance yielded the most consistent results. This procedure allowed us to evaluate \texttt{docExtractor} using the same classification and detection metrics as our pipeline.

\subsection{Image Captioning}
\label{methods-caption}
To enable meaningful retrieval and exploration of the illustrations, we generated a textual caption for each cropped image using a vision-language model. Unlike classification or detection tasks that focus on identifying the presence or location of objects, image captioning addresses the semantic content describing what the image portrays~\cite{vinyals2015show, bernardi2016automatic}. Captions were expected to capture key visual features such as objects, emotions and colors (e.g., “a crowned figure holding a scepter,” “a floral border,” or “a medieval battle scene”).

We evaluated several state-of-the-art models, including Florence-2\cite{florence2}, BLIP \cite{li2022blip}, and LLaVA \cite{li2023llava}, using them off the shelf without additional fine-tuning.
To compare their qualitative performance, we conducted an empirical manual evaluation over a representative set of 100 manuscript illustrations. While we did not apply quantitative metrics (e.g., BLEU or CIDEr) at this stage, the manual inspection revealed noticeable differences in output quality. We observed that LLaVA tended to generate captions that were more accurate, linguistically richer, and better able to reflect the stylistic diversity of the illustrations. Based on this empirical judgment, we selected LLaVA for further development. The generated captions were then stored alongside the associated cropped illustrations and metadata.

\subsection{Search and Retrieval}
\label{methods-search}

In the final stage, the cropped images, their captions, and manuscript metadata were integrated into a searchable database and made accessible through a lightweight web interface. Each record contained the cropped illustration, its automatically generated caption, manuscript identifier, page number, and an IIIF URL linking back to the full digitized volume. By integrating classification, detection, and captioning into a unified modular framework, our system enables scalable semantic search and retrieval of historical visual content. These capabilities provide humanities scholars with direct verbal access to illustrations otherwise hidden within large-scale manuscript collections, opening new possibilities for studying visual culture across time, style, and geography. 

\begin{figure*}[t]
\centering 
\includegraphics[width=0.75\linewidth]{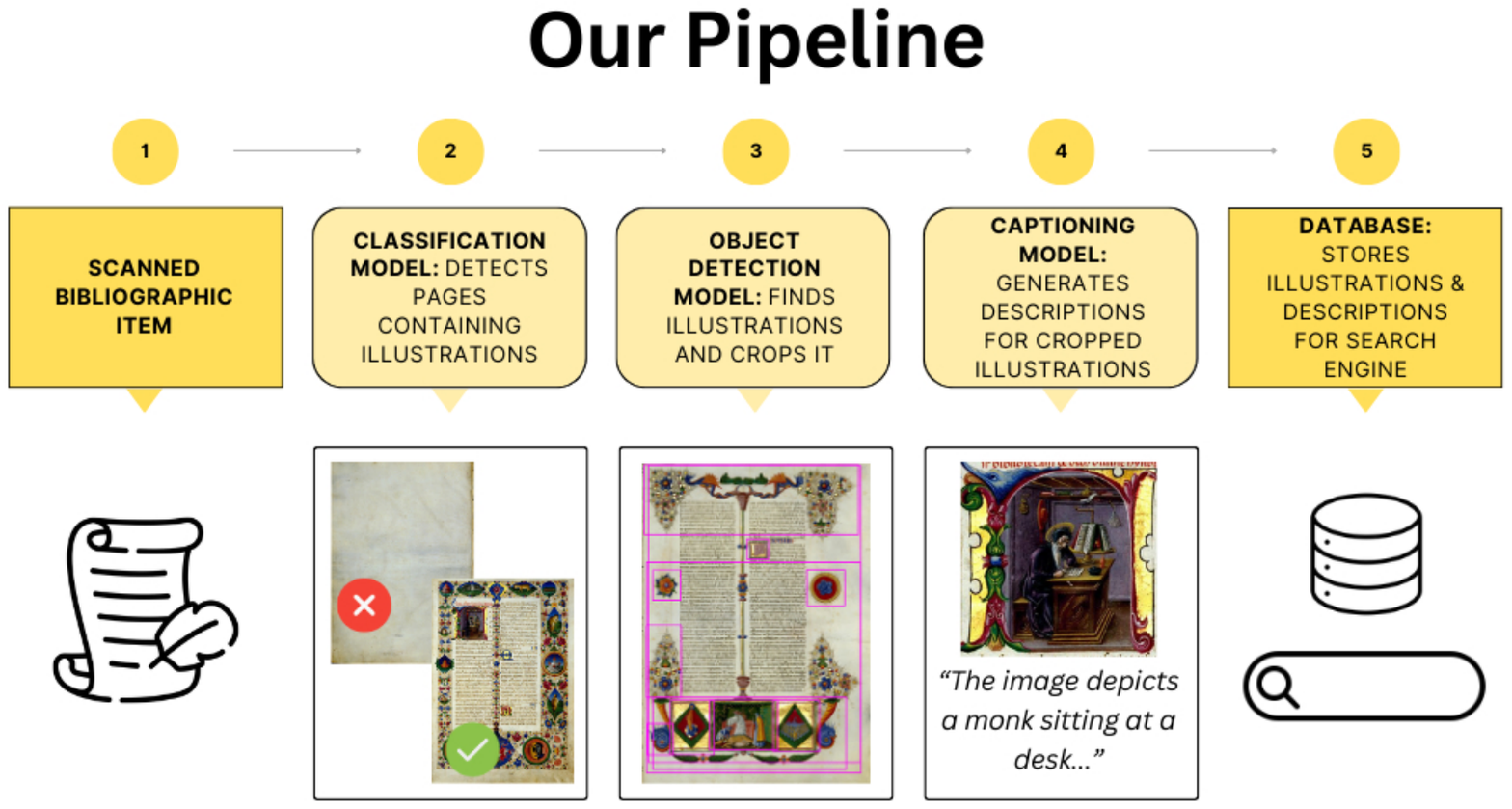}
\caption{Our pipeline for transforming a vast collection of scanned historical document pages into a searchable system for artwork and illustrations.}
\label{fig:image4.png}
\end{figure*}

\subsection{Illustration Similarity Graph}
\label{methods-similarity}
On top of the search and retrieval layer, we generated visual embeddings - numerical representations that capture the stylistic and semantic content of each illustration using CLIP~\cite{radford2021learning}. We measured similarity between pairs of embeddings using cosine similarity and, for each image, connected it to its fifty nearest neighbors in the embedding space. This procedure produced an illustration-similarity graph in which each node corresponds to an extracted image and edges connect visually related pairs.

This graph offers a corpus-level perspective on visual relationships across the collection. Analyzing its structure reveals coherent communities of related images—clusters that reflect shared stylistic features, iconographic motifs, or thematic elements. The graph serves two complementary purposes. First, it enables interactive exploratory browsing: starting from any illustration, scholars can navigate to visually similar images and trace chains of related motifs both within and across manuscripts. Second, applying graph clustering and community detection methods allows the identification of groups of images that share common visual patterns, such as similar compositions, color palettes, or page-layout positions. These communities support higher-level inquiries, for example, tracing the evolution of animal imagery across manuscripts from different periods.


\section{Results}
\label{sec:results}

To evaluate our system’s performance, we tested it on a dedicated dataset of digitized collections from the Vatican Library and the Bible of Borso d’Este.
The Vatican Library offers an exceptionally diverse corpus, spanning a wide temporal range, numerous artistic and scribal traditions, and a rich variety of materials, making it an ideal source for training and evaluating our algorithms on heterogeneous data.
In contrast, Bible of Borso d’Este provide a compact yet illustration-rich item, providing a focused test case that allow us to assess the system’s ability to detect and analyze richly decorated pages within a more constrained corpus.
Our objectives were threefold: (1) accurately distinguishing illustrated pages from text-only ones; (2) detecting and cropping illustrations within pages; and (3) generating meaningful textual descriptions. In this section, we present the results of each stage of the pipeline, followed by an analysis of overall throughput and scalability, including the downstream construction of an illustration similarity graph, which leverages the extracted embeddings to reveal corpus-level visual structure.

\subsection{Classification: Illustration Presence}
We first evaluated our page-level classification model (EfficientNet-B0) on a held-out test set of 1,105 images that were not used during training or validation. The set contained 118 illustrated pages ("art") and 987 text-only pages ("no-art"), mirroring the real distribution while ensuring a balanced evaluation of 10.6\%.

The model demonstrated strong overall discrimination performance, achieving a precision of 78.6\%, recall of 74.6\%,  F1-score of 76.5\%, and an accuracy of 95.1\%. The area under the ROC curve (ROC-AUC) reached 0.95, and the area under the precision-recall curve (PR-AUC) was 0.82, indicating a high ability to separate illustrated from non-illustrated pages despite the natural class imbalance. Notably, when adopting a lower decision threshold (0.2) that prioritizes recall over precision for the "art" class, recall increased to 83\%, capturing a substantially larger share of illustrated pages, at the cost of a modest rise in false positives.

For comparison, a naive baseline that predicts all pages as “non-illustrated” would achieve an apparent accuracy of 89.3\%, reflecting the dominance of text-only pages in the dataset. However, such a model would completely fail to identify any illustrated pages (precision = 0, recall = 0, F1 = 0), providing no practical utility. In contrast, our model achieves both high accuracy (95.1\%) and strong discrimination (precision = 78.6 \%, recall = 74.6 \%, F1 = 76.5 \%).

\subsection{Detection: Illustration Localization}
For illustration localization, the YOLOv11n model was evaluated on a test set of 179 images containing bounding boxes for all visual regions. Results are presented in Table~\ref{tab:detection}.

\begin{table}[H]
\centering
\caption{Illustration detection performance on 179 test images.}
\label{tab:detection}
\resizebox{0.5\textwidth}{!}{%
\begin{tabular}{lcccc}
\toprule
Model & mAP@0.5 & mAP@0.5:0.95 & Precision & Recall \\
\midrule
YOLOv11n & 75.6\%& 51.2\%& 55.3\%& 78.7\%\\
\bottomrule
\end{tabular}
}
\end{table}

Our test dataset consisted of 179 images containing 404 manually labeled illustrations. Of these, 318 were successfully detected by our fine-tuned model, while 86 were missed, yielding a relatively high recall of 0.79. In addition, our model identified 257 illustrations that had not been manually labeled as such, resulting in a precision of 0.55 - meaning that slightly more than half of the detected illustrations were indeed annotated ground-truth illustrations.

\begin{figure}
\centering
\includegraphics[width=\linewidth]{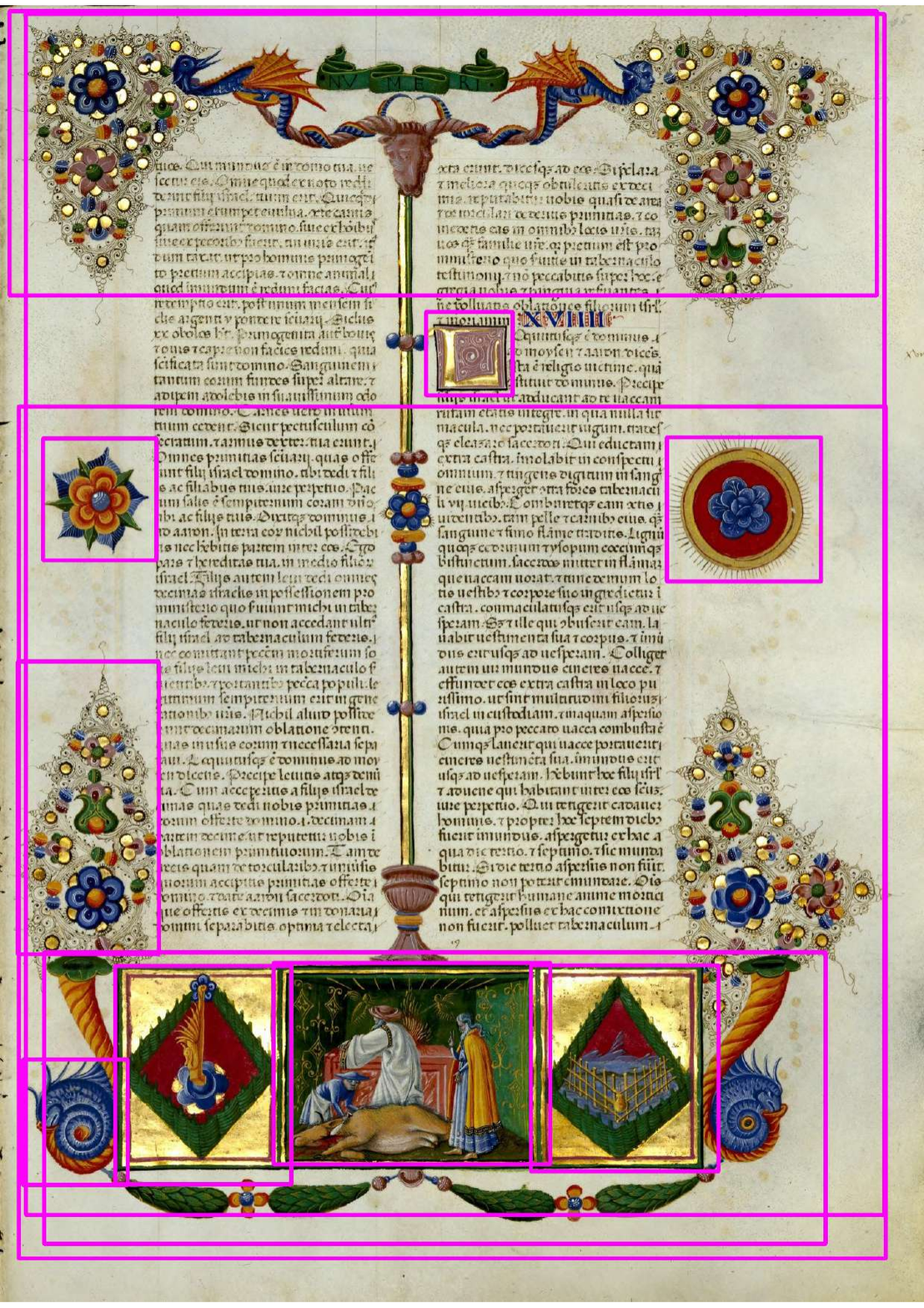}
\caption{Examples of detected illustrations using our fine-tuned YOLOv11n. Bounding boxes highlight illustration regions.}
\label{fig:detection}
\end{figure}

Through the comparison with \texttt{docExtractor}, a segmentation-based approach, their system detected 134 of the 404 ground-truth illustrations while producing 608 false positives. In terms of efficiency, our model processed each page in \textbf{0.06 seconds}, compared with an average of \textbf{51 seconds} for \texttt{docExtractor} on the same hardware. These measurements refer only to the segmentation-mask stage and exclude pre- and post-processing overhead.

\subsection{Captioning: Generating Descriptions}
After illustrations were cropped, we applied LLaVA for caption generation. A qualitative evaluation of 100 randomly sampled illustrations showed that more than 75\% of generated captions accurately described the illustrations and captured the main semantic elements (objects, figures, actions). However, the model struggled with detailed illustrations, sometimes missing finer elements. It also tended to misinterpret decorative initials and occasionally identified abstract shapes as animals or people.

Nevertheless, in most cases the generated captions were sufficiently accurate to enable meaningful keyword-based retrieval. Future improvements could include fine-tuning captioning models on manuscript-specific data to reduce misinterpretations, and as well as incorporating newer models as they become available.

\subsection{System Throughput and Scalability}
To assess scalability, we applied the full pipeline to over five million manuscript pages from the Vatican Library and the Bible of Borso d’Este. The classification stage filtered out more than 90\% of pages as text-only, leaving around 250,000 illustrated pages for further analysis. The detection stage localized around 350,000 distinct illustrations, which were then captioned and indexed in a searchable database.

Overall throughput averaged under 0.06 seconds per page, enabling complete processing of the Vatican corpus in several days on a single GPU server. By contrast, traditional segmentation approaches due to their longer runtime, would require several months or even more for the same corpus.

\begin{figure*}
\centering 
\includegraphics[width=0.75\linewidth]{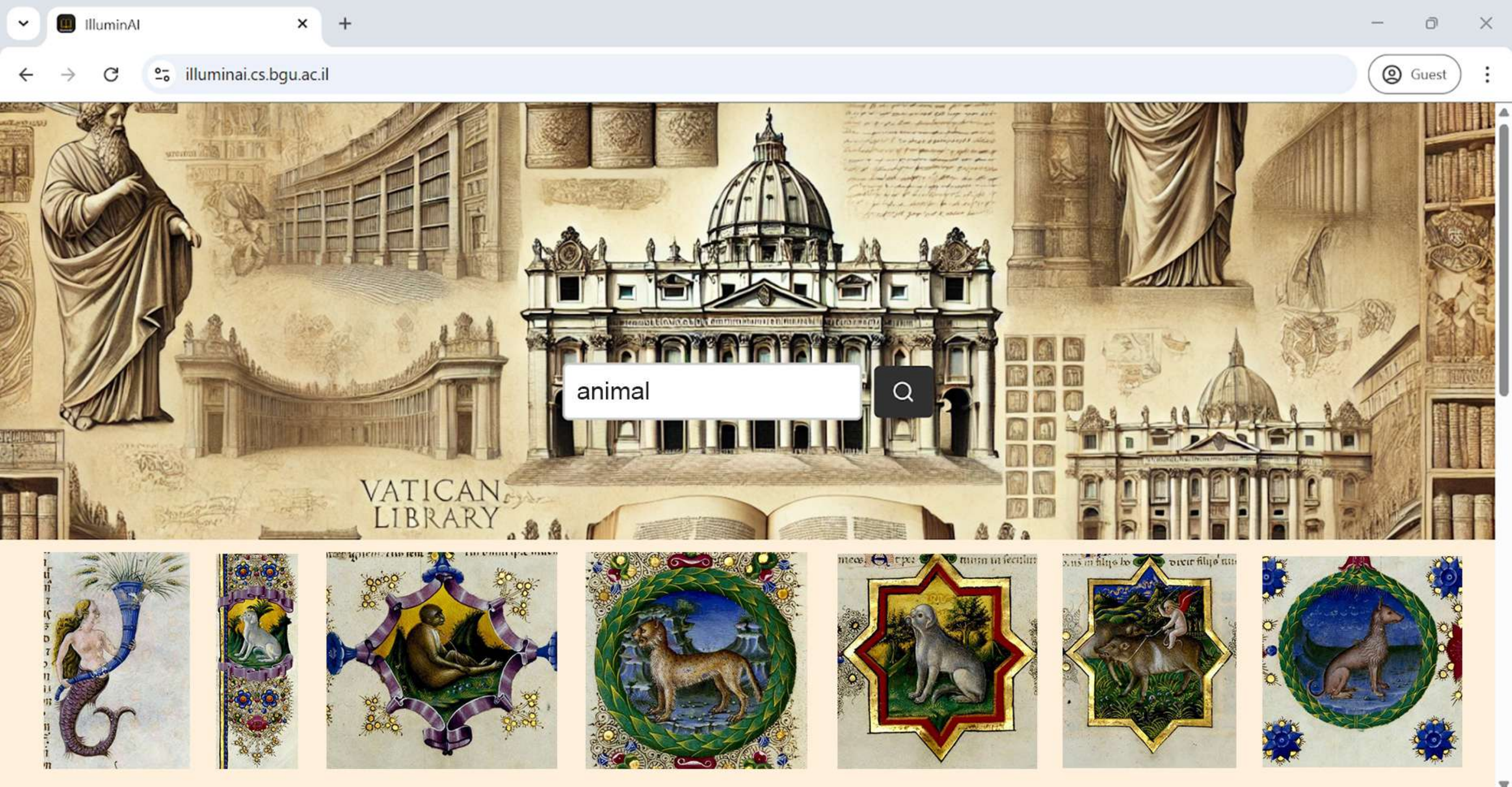}
\caption{An illustration of an example search using our prototype interface on the Bible of Borso.}
\label{fig:search}
\end{figure*}

\subsection{Illustration Similarity Graph}
To demonstrate how the framework can be used not only to retrieve individual illustrations but also to analyse relationships between them, we applied the full pipeline to the Vatican Library and the Bible of Borso d’Este. After running the classification and detection stages on all pages, we automatically extracted the manuscript’s miniatures, historiated initials, and decorative borders. Each cropped illustration was then embedded into a shared visual feature space and linked to its nearest neighbours, yielding an illustration-similarity graph reflects the visual structure of the corpus.

Qualitative inspection of the resulting graph revealed coherent communities of related images. In the case of the Vatican Library’s digitized collection, this method surfaced distinct visual groupings: animals, ornate initials, heraldic symbols, and full-page miniatures naturally clustered together, despite being drawn from different manuscripts, time periods, and artistic traditions. This demonstrates how our framework unlocks visual structures and cross‑manuscript relationships that were previously beyond the reach of manual scholarship, pointing toward a new generation of large‑scale, data‑driven humanities research.
Similarly, in the Bible of Borso d’Este, a richly illuminated fifteenth-century manuscript, our system identified meaningful visual communities, including clusters of decorative borders, historiated initials, and full-page illustrations.

Taken together, these findings show that the illustration similarity graph provides a complementary, network-based perspective on manuscript collections. Instead of viewing pages one at a time, scholars can explore a corpus as an interconnected network of related motifs and visual themes. This enables new research questions about how motifs are repeated, varied, or recombined across a manuscript; how stylistic patterns propagate across workshops or hands; and how visual themes are distributed within or across codices. By uncovering corpus-level visual structure, the graph opens the door to a new generation of large-scale scholarship.

\begin{figure*}[t!]
    \centering
    \begin{subfigure}[t]{\textwidth}
        \centering
        \includegraphics[width=\textwidth]{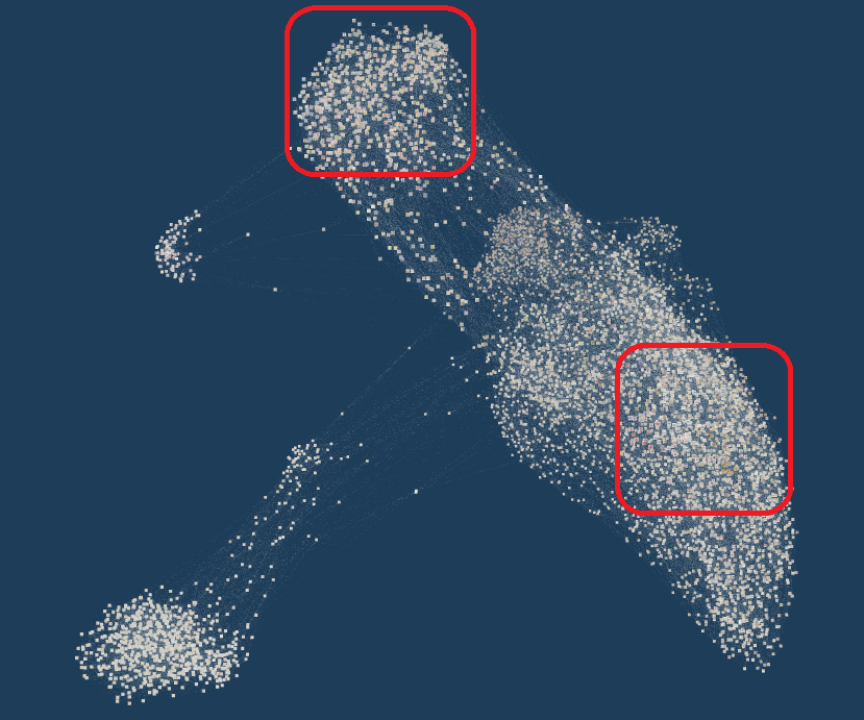}
        \caption{Global illustration similarity network with highlighted regions corresponding to two selected communities.}
        \label{fig:network-marked}
    \end{subfigure}

    \vspace{0.8em}

    \begin{subfigure}[t]{0.48\textwidth}
        \centering
        \includegraphics[width=\textwidth]{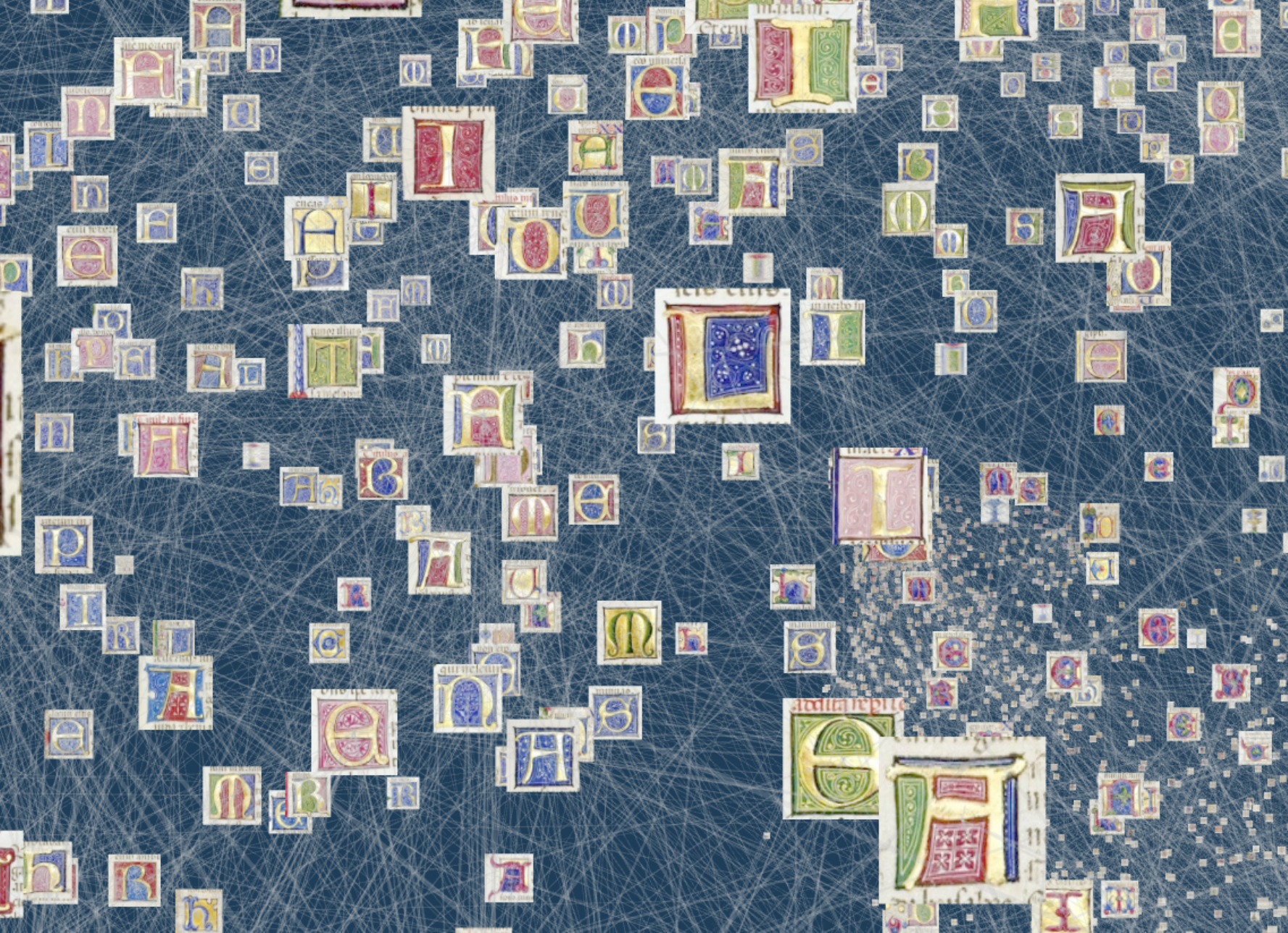}
        \caption{Zoomed-in view of a community dominated by decorated initials.}
        \label{fig:network-chars}
    \end{subfigure}
    \hfill
    \begin{subfigure}[t]{0.48\textwidth}
        \centering
        \includegraphics[width=\textwidth]{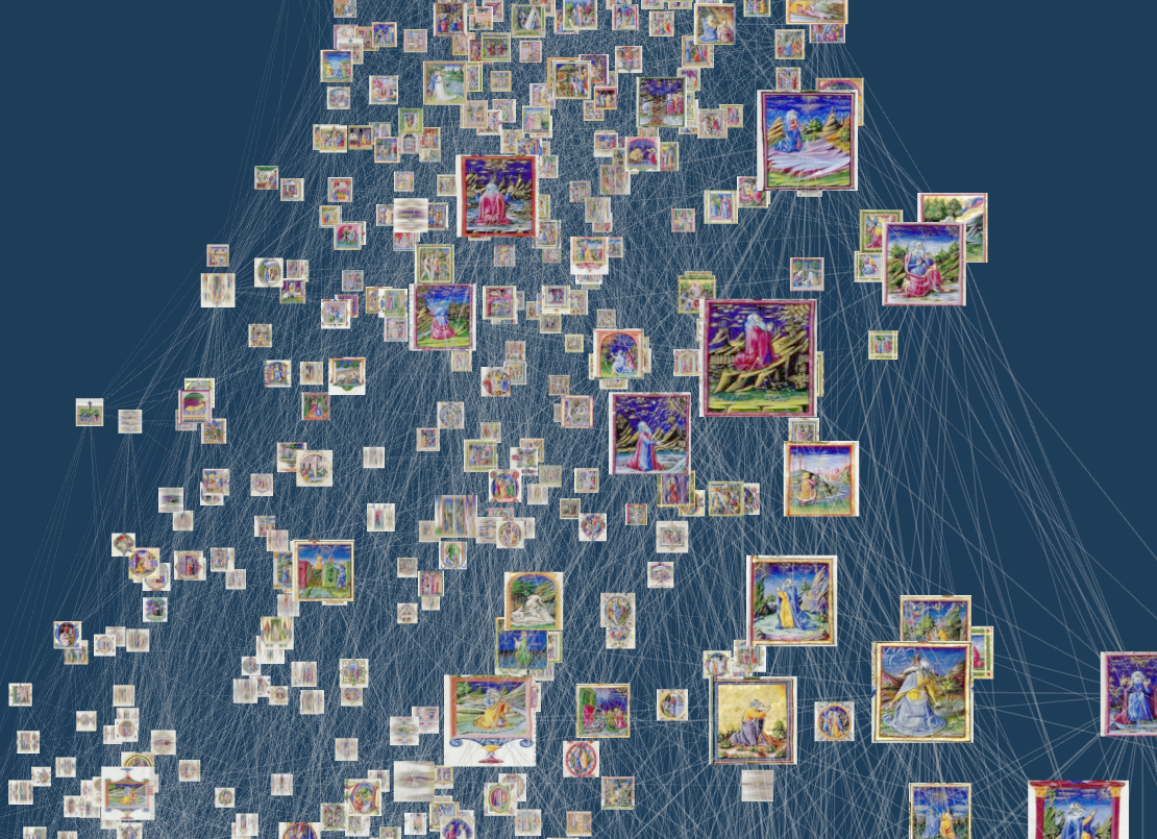}
        \caption{Zoomed-in view of a community composed of general painted scenes.}
        \label{fig:network-drawing}
    \end{subfigure}

    \caption{
    Similarity network of illustrations from the Bible of Borso d’Este.
    (a) The full network layout, where two regions are marked to indicate communities selected for closer inspection.
    (b) A zoomed-in cluster centered on decorated initials.
    (c) A zoomed-in cluster consisting primarily of narrative and ornamental painted scenes.
    }
    \label{fig:network}
\end{figure*}


\section{Discussion} 
\label{sec:discussion}
Efficiently locating illustrated pages within large-scale digitized manuscript collections can profoundly advance research in art history and the digital humanities. Our study introduces a scalable deep-learning framework capable of automatically identifying, extracting, and describing illustrations across heterogeneous historical corpora. To achieve this goals, our work was structured into several stages: 

First, we fine-tuned a deep learning image classification model to distinguish between illustrated and non-illustrated pages. Its strong performance across multiple metrics including AUC, precision, recall, accuracy, and F1-score demonstrates the model’s overall effectiveness. Although roughly  one quarter of illustrated pages were missed, yet the vast majority of irrelevant text-only pages were correctly excluded. Since illustrated pages represent only a small fraction of the corpus, minimizing false positives is crucial; even a modest false-positive rate could otherwise overwhelm the system with millions of irrelevant results. By successfully filtering these pages, the model ensures that users encounter focused, high-quality outputs, thus enhancing the system’s overall precision and usability. Future improvements could be achieved by addressing class imbalance more effectively, and expanding the training dataset.

Next, we applied an object detection model to locate and crop individual illustrations within each identified page. Its high recall indicates that the majority of illustrations were successfully captured, though some were occasionally missed. Upon examining all such cases  we observed two consistent patterns: (1) initial letters that were not highly ornamented and thus resembled regular text, and (2) illustrations composed of multiple “sub-illustrations.”  
To investigate the model’s moderate precision (0.55), we analyzed the false positives - regions detected as illustrations by the model but not labeled as such manually. A consistent trend emerged: nearly all of these were sub-illustrations (e.g., characters, animals, or other elements within a larger drawing) that had not been annotated during manual labeling. Notably, we found almost no false positives in which stains, text, or irrelevant visual artifacts were misclassified as illustrations.
Although our object detection model correctly identifies nearly 80 \% of illustrations, we expect that stricter annotation guidelines, particularly regarding sub-illustrations and larger training datasets will further improve performance and reduce both false negatives and false positives at this stage.

The comparison we conducted between our system and \textit{docExtractor} revealed several noteworthy patterns. Upon reviewing their outputs, we observed that the model segmented illustration pixels with high accuracy. However, the pre- and post-processing steps required to convert pixel-level masks into bounding boxes made it difficult to precisely delineate illustration borders. This issue was especially evident on pages containing multiple or composite illustrations, where morphological closing often merged nearby elements into single regions, leading to missed detections or fragmented false positives. Similarly, many of the apparent false positives corresponded to fragments of valid illustrations that could not be cleanly separated  as independent objects.

We believe that this challenge highlights the difficulty of adapting segmentation outputs for retrieval-oriented pipelines. Additionally, it is worth noting that a key advantage of our system lies in its speed, making it particularly suitable for large-scale corpora. 
These observations underscore the methodological gap between segmentation and object detection when applied at scale to vast digital repositories, especially in retrieval scenarios where sub-illustration recovery is crucial.  By shifting from pixel-level segmentation to page-level classification and object detection, we dramatically reduce computation time while still providing reliable identification of images likely to contain artistic or decorative elements.

Another significant challenge is the diversity of historical manuscripts. Although we carefully balanced our dataset and ensured robust training, the true range of manuscript traditions extends far beyond the Vatican Library and the Bible of Borso d’Este used in our case study. Manuscripts from different cultural traditions, regions, and time periods often exhibit distinct styles, materials, and visual conventions. Validating our model on digitized collections from other libraries and cultures will therefore be essential to strengthen robustness and ensure generalizability across varied datasets.

In the third stage, we employed LLaVA as an off-the-shelf image captioning model. While our current evaluation of image-captioning models was limited to an empirical qualitative comparison, future work should include a more systematic investigation of captioning performance across a wider range of architectures. This could involve applying established quantitative metrics (e.g., BLEU, CIDEr) and expanding the benchmark to include additional, more recent vision-language models such as Qwen-VL~\cite{bai2023qwenvlversatilevisionlanguagemodel} and Llama 4 Vision~\cite{meta2025llama4}, which may improve descriptive accuracy and richness. ~~Also,~~ Future fine-tuning on this specific domain could further enhance the performance of these models for manuscript illustration tasks.

In addition to the three core stages of our pipeline, we constructed an illustration similarity graph to evaluate how well the extracted illustrations and their embeddings capture higher-level visual relationships. This step allowed us to assess whether the embedding space meaningfully organizes the corpus beyond simple page level retrieval. Our qualitative inspection showed that the embeddings grouped together illustrations with consistent motifs and styles. In both the Vatican Library collection and the Bible of Borso d’Este, we observed coherent communities, for example, clusters of border decorations, designed initials, animals, and botanical illustrations. Notably, many of these relationships spanned distant folios or even different manuscripts, revealing connections that are difficult to notice when browsing collections page by page.

While these clusters looked promising, more systematic evaluation is still required. Because the embedding model was trained on modern image datasets rather than manuscript art, future work should include additional validation to ensure that visually or semantically meaningful relationships are consistently captured.

Taken together, the graph analysis provides a complementary perspective to the detection and captioning stages. This layer enables scholars to explore cross-manuscript “neighborhoods” of related illustrations and visual patterns that are often invisible at the page level. Also, the graph offers a scalable and efficient way to study how artistic themes emerge, evolve, and recur across manuscripts, traditions, and historical periods. For example, researchers could examine how depictions of animals change across centuries or how ornamental border styles propagate across different workshops and geographic regions.

In general, it is important to emphasize that our pipeline is fully modular. In this study, we selected specific algorithms for each stage - page classification, illustration detection, and illustration captioning. However, as new architectures continue to emerge, each component of the pipeline can be seamlessly replaced or upgraded to incorporate more advanced models in future work.

From a digital humanities perspective, the ability to quickly isolate illustrated pages from the vast sea of manuscripts offers new and groundbreaking avenues for research. Scholars in Jewish Studies may, for example, focus on illuminations tied to specific scriptural passages or rabbinic commentaries, thereby reconstructing the transmission of visual motifs across geography and time. 
Similarly, scholars of Christian art and theology may investigate how biblical scenes and hagiographic motifs were visually interpreted in illuminated manuscripts, tracing the diffusion of stylistic and devotional traditions across monasteries and scriptoria in Europe.
Likewise, art historians can prioritize the digitized pages most likely to yield insights into medieval painting and manuscript decoration, saving significant effort in the face of overwhelming digital data.

By efficiently identifying and categorizing illustrated pages, researchers can detect trends across time periods and regions, revealing shifts in artistic styles, cultural exchanges, and iconographic traditions. This also enables systematic study of the relationship between text and image, highlighting how visual content complemented written narratives. In addition, it supports large-scale comparative analysis, allowing similarities and differences between collections, regions, or time periods to be examined with greater precision.

Finally, despite its promising results, our approach has several limitations. First, as with many deep-learning systems, the quality and diversity of the training data heavily influence performance. The model may struggle with rare visual styles, unusual color palettes, or damaged pages that differ significantly from the training examples. Second, manual annotations, particularly for complex or partially decorated pages are inherently subjective, introducing noise that may affect both classification and detection accuracy. They also depend on contextual knowledge and cultural associations that are not always accessible to an artificial intelligence model, or that require significant exposure and adaptation time to be effectively learned. Moreover, such tools can also empower researchers in fields like zoology, botany, or environmental history to systematically identify depictions of animals and plants within manuscripts and study how different species were represented, symbolized, and understood in various cultural and historical contexts.

In addition, the image captioning stage remains imperfect: the model occasionally misidentifies well known historical or religious figures, describing, for instance, Jesus simply as “a man.” Such oversights, which are immediately recognizable to human experts, can influence the precision of the retrieval system. Moreover, current captioning models tend to overlook subtle visual cues such as emotions, gestures, or symbolic details that are essential for nuanced interpretation in art-historical research.


\section{Conclusions}
\label{sec:conclusions}

Illustrations in historical manuscripts provide critical evidence for understanding artistic practices, cultural exchange, and intellectual traditions. With the rapid growth of digitized collections, however, traditional document analysis methods such as page layout analysis or pixel-level segmentation are increasingly insufficient for large-scale exploration.

In this work, we introduced a deep-learning pipeline for large-scale illustration extraction, integrating image classification (EfficientNet) and object detection (YOLO) fine-tuned on datasets from the Vatican Library and the Bible of Borso d’Este. The system detects illustrated pages and localizes drawings with inference times below 0.06 seconds per page. Beyond detection, we incorporate an image-language model that generates textual captions, enabling keyword-based search and retrieval across millions of pages, and ultimately support an illustration-similarity graph that reveals visual relationships and recurring motifs across manuscripts.

Although the system does not detect every illustration, it successfully identifies the vast majority, substantially reducing the reliance on manual labor and enabling access to repositories at scales that were previously impractical. This capability opens opportunities for cross-collection comparisons, iconographic studies, and the integration of visual evidence into broader interdisciplinary research.

Looking ahead, we plan to improve the accuracy of the classification and detection models. Our robust, user-friendly search engine thus lays the groundwork for exploring visual culture across diverse periods and regions, contributing to a deeper understanding of artistic practices, cultural exchange, and the transmission of visual traditions over time.


\section{Code and Data Availability} 

All code, data, and trained models associated with this research will be made publicly available upon publication of the paper.


\section*{Acknowledgments} 

We gratefully acknowledge the generous support of The Soref-Breslauer Texas Foundation, whose contribution made this research possible.

While drafting this article, we used ChatGPT~\cite{openai2023chatgpt} for English editing to enhance the clarity and richness of the text in this paper.


\phantomsection

\bibliographystyle{IEEEtran}
\bibliography{citations}


\end{document}